\documentclass{pasa}%
\usepackage{morefloats}
\title[Local stellar kinematics from RAVE data - VI.]{Local Stellar Kinematics from RAVE data - VI. Metallicity Gradients Based on the F-G Main-sequence Stars}
\author[Plevne et al.]{O. Plevne$^1$\thanks{olcayplevne@gmail.com}, T. Ak$^2$, S. Karaali$^2$, S. Bilir$^2$, S. Ak$^2$, and Z.~F. Bostanc\i$^2$\\
     \affil{$^1$Istanbul University, Graduate School of Science and Engineering, Department of Astronomy and Space Sciences, 34116, Beyaz\i t, Istanbul, Turkey}
     \affil{$^2$Istanbul University, Faculty of Science, Department of Astronomy and Space Sciences, 34119, Beyaz\i t, Istanbul, Turkey}
}
\jid{PASA}
\doi{10.1017/pas.\the\year.xxx}
\jyear{\the\year}
\usepackage{natbib}
\begin{document}

\begin{abstract}
We estimated iron and metallicity gradients in the radial and vertical directions with the F and G type dwarfs taken from the RAdial Velocity Experiment (RAVE) Data Release 4 (DR4) database. The sample defined by the constraints $Z_{max}\leq 825$ pc and $e_p\leq0.10$ consists of stars with metal abundances and space velocity components agreeable with the thin-disc stars. The radial iron and metallicity gradients estimated for the vertical distance intervals $0<Z_{max}\leq 500$ and $500<Z_{max}\leq 800$ pc are $d[Fe/H]/dR_m=-0.083\pm0.030$ and $d[Fe/H]/dR_m=-0.048\pm0.037$ dex~kpc$^{-1}$; and $d[M/H]/dR_m=-0.063\pm0.011$ and $d[M/H]/dR_m=-0.028\pm0.057$ dex~kpc$^{-1}$, respectively, where $R_m$ is the mean Galactocentric distance. The iron and metallicity gradients for less number of stars at further vertical distances, $800<Z_{max}\leq 1500$ pc, are mostly positive. Compatible iron and metallicity gradients could be estimated with guiding radius ($R_g$) for the same vertical distance intervals $0<Z_{max}\leq 500$ and $500<Z_{max}\leq 800$ pc, i.e. $d[Fe/H]/dR_g=-0.083\pm0.030$ and $d[Fe/H]/dR_g=-0.065\pm0.039$ dex~kpc$^{-1}$; $d[M/H]/dR_g=-0.062\pm0.018$ and $d[M/H]/dR_g=-0.055\pm0.045$ dex~kpc$^{-1}$. F and G type dwarfs on elongated orbits show a complicated radial iron and metallicity gradient distribution in different vertical distance intervals. Significant radial iron and metallicity gradients could be derived neither for the sub-sample stars with $R_m\leq 8$ kpc, nor for the ones at larger distances, $R_m>8$ kpc. The range of the iron and metallicity abundance for the F and G type dwarfs on elongated orbits, [-0.13, -0.01), is similar to the thin-disc stars, while at least half of their space velocity components agree better with those of the thick-disc stars. The vertical iron gradients estimated for the F and G type dwarfs on circular orbits are $d[Fe/H]/dZ_{max}=-0.176\pm0.039$ dex~kpc$^{-1}$ and $d[Fe/H]/dZ_{max}=-0.119\pm0.036$ dex~kpc$^{-1}$ for the intervals $Z_{max}\leq 825$ and $Z_{max}\leq 1500$ pc, respectively. 
\end{abstract}

\maketitle%

\section{Introduction}
Metallicity gradient is one of the tools used for understanding the structure, formation and evolution of the Milky Way Galaxy. Systematic sky surveys, which use photometric, astrometric and spectroscopic observations, are needed to reveal the metallicity gradients of the Milky Way. These surveys are used to determine the atmospheric model parameters and kinematic properties of a large number of objects simultaneously. Specifically, Genova-Copenhagen Survey \citep[GCS;][]{Nordstrom04}, RAdial Velocity Experiment \citep[RAVE;][]{Steinmetz06}, Sloan Extension for Galactic Understanding and Exploration \citep[SEGUE;][]{Yanny09}, the Large Sky Area Multi-Object Fiber Spectroscopic Telescope \citep[LAMOST;][]{Deng2012,Zhao2012}, and the {\it Gaia}-ESO Spectroscopic Survey \citep{Gilmore2012} are among the most useful systematic sky surveys. In order to estimate the metallicity gradient in any direction, distances of the objects in question must be known. Different objects such as F and G type main-sequence stars (dwarfs), red clump stars, open and globular clusters, and cepheids have been used for this purpose, for which the distances can be accurately estimated. The metallicity gradients are estimated either vertically or in the Galactocentric radial direction. The radial metallicity gradients, $d[M/H]/dR$, in different vertical distance ($z$) intervals from the Galactic plane have been presented in the previous studies. A metallicity gradient often indicates the inward-outside formation of the Galactic part in question, such as the thin disc \citep{Matteucci89, Chiappini97, Cescutti07, Pilkington12}. However, \cite{Schonrich09} showed that an inward-outside formation is not necessary for a metallicity gradient. 

The vertical and radial distances of a star sample can be estimated for its observed position, or they can be obtained from its possible Galactic orbit as determined using the Milky Way potential (Section 2). Usually, we use the $z$ and $R$ symbols for the observed vertical and radial distances, respectively, while $Z_{max}$, $Z_{min}$, and $R_{m}$ are the maximum and minimum distances of a star to the Galactic plane, and the mean Galactocentric radial distance, respectively. $R_m$ is defined as $R_m=(R_p+R_a)/2$, where $R_p$ and $R_a$ are the perigalactic and apogalactic distances, respectively. The scale of the $R$ distances for the objects such as giants can be used in the metallicity gradient evaluations. However, this parameter as well as $Z$ may not represent the birth place of a star, due to its large orbital eccentricity. For dwarfs, the orbital parameters $Z_{max}$ and $R_m$ had to be used in order to extend the scale in the vertical and radial distances. However, in this case, we can not avoid a bias as explained below: Only stars with large orbital eccentricities among a sample of small $R_m$ distances can reach to the solar circle. But, such stars are metal-poor relative to the ones on the circular orbits with the same $R_m$ radial distances which can not reach to the solar circle. Thus, the number of the metal-rich stars at the solar circle decreases. This is a bias against metal-rich stars with small $R_m$ radial distances \citep{Boeche14}.

The researchers found iron metallicities in the range $-0.17\leq d[Fe/H]/dR\leq -0.06$ dex~kpc$^{-1}$ for the Galactic thin disc. The value at the left is estimated by \citet{Sestito08} with open clusters, while the right one -corresponds to the cepheids- appeared in \citet{Luck11}. An iron metallicity gradient close to $d[Fe/H]/dR=-0.06$ dex~kpc$^{-1}$ can be found in recent studies which are based on the data from large surveys such as RAVE and SEGUE, i.e. \citet{Cheng12} and \citet{Boeche13, Boeche14}. The radial metallicity gradients in these recent studies are given in different $Z_{max}$ intervals. In these studies, the radial metallicity gradient becomes less negative with increasing $Z_{max}$ and approach to zero or even to a positive value at the terminal vertical distance. It is assumed that the different numerical values of the radial metallicity gradient at large vertical distances are due to the different structure of the thick disc; i.e. the metallicity gradient of the thick disc is almost zero, if it exists \citep{Ruchti11, Cheng12, Coskunoglu12, Bilir12}. Stars at different ages exhibit substantial differences in radial metallicity gradient \citep{Schonrich09, Minchev13, Minchev14}; hence, investigation of the metallicity gradients in narrow $Z_{max}$ intervals plays a limiting role on these differences. Actually, thin-disc stars concentrate close to the Galactic mid-plane as compared to the farther thick-disc stars, which are older than the thin-disc stars. We wish to refer the recent results on gradients from the SDSS APOGEE, as reported in \citet{Hayden14} and \citet{Anders14}. \citet{Hayden14} observed radial gradients as a function of $|z|$ for a variation of subsamples including low-$\alpha$ and high-$\alpha$ stars, while \citet{Anders14} give a metallicity gradient of $d[M/H]/dR=-0.082\pm0.002$ dex~kpc$^{-1}$ for their sample. 

As stated above, different objects, which represent the structure of the interstellar gas at the time they were formed, are used in metallicity estimations. These  objects therefore correspond to different metallicity gradients. However, the situation is not so simple as explained in the following text. Some processes such as the spiral structure, molecular clouds or small aggregations causes a dynamical heating of the Galaxy which change the orbits of the objects in question \citep[cf.][]{Cheng12}. Radial mixing is another constraint which changes the original positions of the objects used in metallicity gradient estimations. The main processes related this constraint are the irregularities in the Galactic potential \citep{Wielen77, Wielen96, Haywood08} and the passage of spiral patterns \citep{Sellwood02, Desimone04, Minchev06}. As a result of the radial mixing, super metallicity stars ($[Fe/H]>0.2$ dex) migrate from the inner part of the disc to the solar circle, while the metal-poor ones ($-0.7<[Fe/H]<-0.3$ dex) change their formation place outside of the disc to the solar circle. Additionally, details about this constraint can be found in \cite{Haywood08}. 

There are substantial differences between the kinematic and spectroscopic parameters of thin and thick discs. Although the observed parameters for the two types of discs are local records and do not cover the regions at remote distances, these differences have been used by researchers as clues to propose a different formation history for the thick disc. The zero (or close to this value) radial metallicity gradient for the thick disc is also an indication of a different formation scenario for the thick disc. 

In this study, our main purpose is to investigate the metallicity gradients of the thin disc in two directions; i.e. vertically and in the Galactocentric direction. Two constraints are used to construct a thin-disc star sample: $i$) we restrict the vertical distances of the sample stars with $Z_{max}\leq 825$ pc, and $ii$) we use only the stars with planar eccentricities $e_p\leq 0.10$. The value $Z_{max}=825$ pc corresponds to the vertical distance where the space densities for the thin and thick discs are compatible, and by adopting a planar eccentricity less than 0.10 we aim to base our estimation on the stars at the solar circle. Thus, we expect less contamination from the thick-disc stars and from the stars migrated from the inner or outward regions of the disc. Eccentricity plays an effective role in discrimination of the thin and thick discs; in addition, it may vary within either of these populations with age. We investigated the metallicity gradient for stars on elongated orbits as well for comparison purposes. The paper is designed follows: Section 2 is devoted to the data, while the results and discussion are presented in Sections 3 and 4, respectively. 

\section{The Data}
\subsection{The Sample}
The sample consists of F and G type dwarfs selected from the recent RAVE Data Release 4 (DR4) database \citep{Kordopatis13} which are identified as explained. The temperature scale is limited to $5310\leq T_{eff}$(K)$\leq 7310$ \citep{Cox00}. The surface gravities of the star sample are identified by the zero age main sequence (ZAMS) and the terminal age main sequence (TAMS) lines via mass tracks in the $\log T_{eff}$-$\log g$ plane as taken from \citet{Ekstrom12}, who adopted the solar abundance as $Z_{\odot}=0.014$ (Fig. 1). We assumed that stars satisfying the stated conditions are F and G type dwarfs. We omitted the stars with $|b|\leq10^\circ$ to avoid high extinction. We used the $E(B-V)$ colour excesses taken from \cite{Schletal11} using IPAC\footnote{http://irsa.ipac.caltech.edu/applications/DUST/}, which are more reliable than the RAVE ones for de-reddening the colours and magnitudes of the sample stars. The near-infrared ($JHK_{s}$) magnitudes of the Two-Micron All Sky Survey \citep[2MASS;][]{Skrutskie06} were taken from the 2MASS All-Sky Catalog of Point Sources \citep{Cutri03}. Next, we used the procedure in \citet{Bilir08} to estimate their $M_J$ absolute magnitudes, which provide distances in combination with the true apparent $J_0$ magnitudes. The $(J-H)_0$ vs $M_J$ colour-absolute magnitude diagram of the star sample (15373 stars in total) thus obtained is given in Fig. 2. The range and the median value of the colour excess for the star sample is $0\leq E(B-V)\leq 0.25$ and $E(B-V)=0.023$ mag, respectively. The errors estimated for the apparent -and absolute- magnitudes are $\Delta J\leq \pm 0.04$ and $\Delta M_J=\pm 0.19$ mag, respectively, which result an error of $\Delta \mu\sim \pm0.20$ mag in $(J-M_J)_o$ distance modulus.  

\begin{figure}
\centering
\includegraphics[scale=0.35, angle=0]{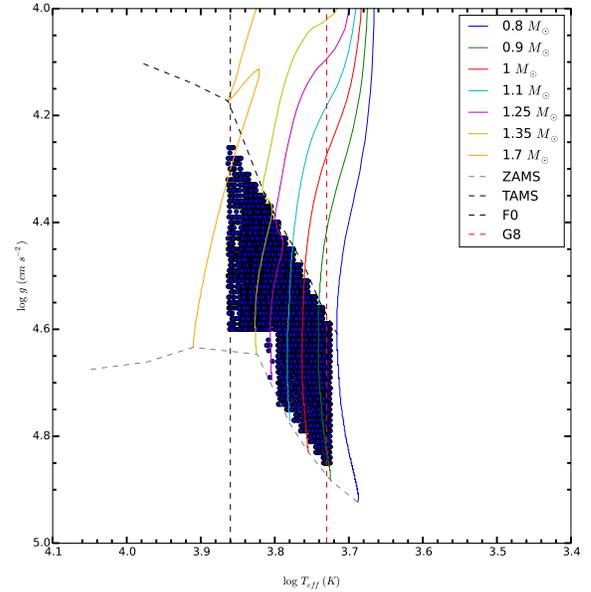}
\caption{The $\log T_{eff}$-$\log g$ diagram for the F and G type dwarfs identified by means of the mass tracks of \citet{Ekstrom12}. Lower and upper bounds indicate the positions of ZAMS and TAMS.} 
\end {figure} 

\begin{figure}
\centering
\includegraphics[scale=0.35, angle=0]{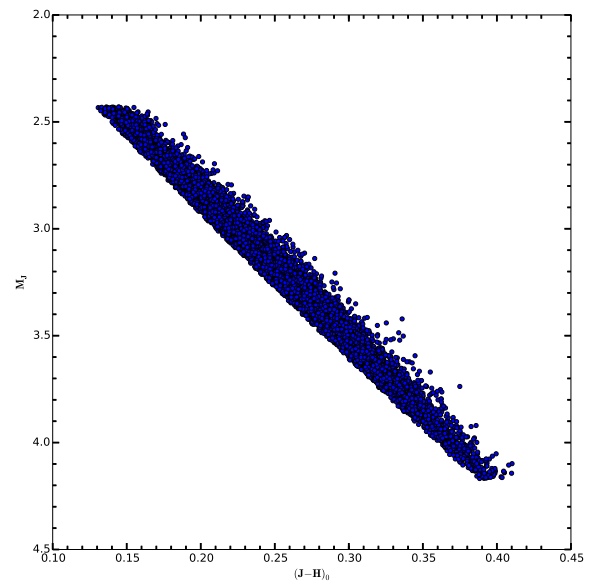}
\caption{$(J-H)_0$ vs $M_J$ colour-absolute magnitude diagram of 15373 F and G type dwarfs identified by means of the mass tracks of \citet{Ekstrom12}.} 
\end {figure} 

The radial velocities ($\gamma$) of the sample stars are taken from the RAVE DR4 database. The radial velocities of about three dozen  stars with an $SNR\sim$14 (signal-to-noise ratio) are rather large; hence, these stars were discarded from the sample. In addition, we also removed the stars beyond the $\pm 3\sigma$ (i.e. $|\gamma|>115$ km~s$^{-1}$) limit of the radial velocities. Thus, the number of the sample stars was reduced to 14927 with $SNR\geq14$. The proper motions of the sample stars are those published in the RAVE DR4  database which contains the data in Tycho-2 \citep{Hog00}, UCAC2, UCAC3, UCAC4 \citep{Zacharias04, Zacharias10, Zacharias13}, PPMX, PPMXL \citep{Roeser08, Roeser10}, and SPM4 \citep{Girard11}.

The temperature scale of the sample stars, F and G dwarfs, is limited to $5310\leq T_{eff}(K)\leq 7310$ \citep{Cox00} and their surface gravities are identified by ZAMS and TAMS lines via mass tracks in the $\log T_{eff}-\log g$ plane, taken from \citet{Ekstrom12}. We tested any probable bias in metallicity by comparing the metallicity distribution for the stars in our sample with the distribution  for all dwarf stars in the effective temperature range $5310\leq T_{eff}(K)\leq 7310$ and surface gravity $4\leq \log g\leq 5$. Our sample consists of 14927 dwarfs, while the number of stars in the second sample, which are identified without any restriction for mass tracks, is 30500. Our research indicated that the metallicity distributions in the two samples are Gaussian with compatible medians, $[Fe/H]= -0.04$ dex and $[Fe/H]=-0.02$ dex for our sample and for the second one, respectively, and their standard deviations are exactly equal at $\sigma=0.25$ dex. Based on these results, we believe that there is no any significant metallicity bias in our sample. 

Another argument that our sample is metallicity un-biased is that metallicity is weakly temperature dependent. Since distances of stars in our data set are not more than 800 pc (see Section 2.2), we conclude that the RAVE DR4 data in this study include the thin disc main-sequence stars in the solar vicinity. Thus the number of metal poor stars in the sample is negligible.

\subsection{Space Velocities and Galactic Orbits}
We combined the distances, radial velocities and proper motions detailed in Section 2.1 and applied the standard algorithms and transformation matrices of \citet{Johnson87} to obtain their Galactic space velocity components, $U$, $V$ and $W$. The procedure is explained in detail in \citet{Coskunoglu11, Coskunoglu12}. Here we will give only the results we obtained. Correction for differential Galactic rotation is necessary for accurate determination of $U$, $V$ and $W$ velocity components. We applied the procedure of \citet{Mihalas81} to the distribution of the sample stars and estimated the first-order Galactic differential corrections for the $U$ and $V$ velocity components of the sample stars. The velocity component $W$ is not affected by Galactic differential rotation \citep{Mihalas81}. The range of the corrections is $-19.28\leq dU\leq 10.95$ and $-1.25\leq dV\leq1.58$ km~s$^{-1}$ for $U$ and $V$ space velocity components, respectively.

The uncertainties in the space-velocity components $U_{err}$, $V_{err}$ and $W_{err}$ were computed by propagating the uncertainties in the proper motions, distances and radial velocities via an algorithm described by \citet{Johnson87}. The median values and the standard deviations for the errors of the velocity components are (\~U$_{err}$, \~V$_{err}$, \~W$_{err}$)=($1.82\pm0.98$, $1.85\pm0.99$, $1.82\pm0.86$) km s$^{-1}$. Next, the error for the total space motion of a star follows from the equation: 

\begin{equation}
S_{err}^{2}=U_{err}^{2}+V_{err}^{2}+W_{err}^{2}.
\end{equation} 
The median value and the standard deviation of the total error are \~S$_{err}=3.37$ and $\sigma=2.67$ km s$^{-1}$, respectively. 

We applied a final constraint to obtain to the space velocity error to determine the stars with high precision. We rejected stars with errors beyond median value plus $2\sigma$ of the total space velocity error, \~S$_{err}\geq 8.71$ km~s$^{-1}$, reducing the sample to 14361 stars which is 96\% of the former sample size. The median values of the rectangular solar centred coordinates of the sample stars are $X=Y=Z=0$ pc. Although the distances of the sample stars lie up to about 900 pc, the number of distant stars is small and the median of the whole sample is only 276 pc (Fig. 3).         

\begin{figure}
\centering
\includegraphics[scale=0.35, angle=0]{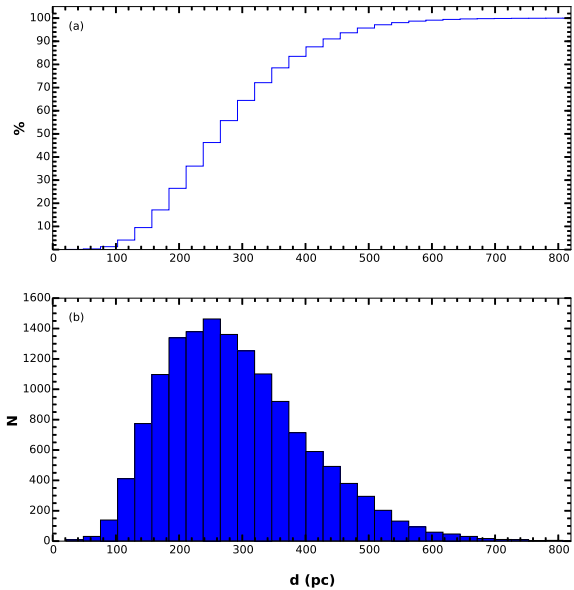}
\caption{Distance histogram for the final 14361 dwarfs where a) shows cumulative sample and b) the number of stars.} 
\end {figure} 

To determine a possible orbit for a given sample star, we performed test-particle integration in a Milky Way potential consisting of three components: halo, disc and bulge potentials. We used {\it MWpotential2014} code of \cite{Bovy15} to calculate the Galactic orbit parameters of the sample stars. The parameters and properties of the Galactic components were taken from the Table 1 of \cite{Bovy15}. The orbital parameters used in this study are $Z_{max}$: maximum distance to the Galactic plane, $R_{a}$ and $R_{p}$: apogalactic and perigalactic radial distances, $R_{m}$: the arithmetic mean of $R_{a}$ and $R_{p}$, $R_{m}=(R_a+R_p)/2$, and $e_p$: the planar eccentricity defined as $e_p=(R_a-R_p)/(R_a+R_p)$. The orbital parameters of the sample stars were calculated within the integration time of 3 Gyr. This integration time corresponds to about 15 revolutions around the Galactic centre so that the average orbital parameters can be determined reliably.

\begin{table*}[!htb]
\centering
\caption{Radial iron and metallicity gradients for the F and G type dwarfs on circular orbits, $e_p\leq0.10$, for different $Z_{max}$ intervals. $N$ indicates the number of stars. Signal to noise ratio is $S/N\geq40$.}  
\begin{tabular}{ccccccc}
\hline
$Z_{max}$ range&$<[Fe/H]>$&$d[Fe/H]/dR_m$ &$N$&$<[M/H]>$&$d[M/H]/dR_m$ &$N$\\
\hline
      (pc)&(dex)&(dex~kpc$^{-1}$) & &(dex)& (dex~kpc$^{-1}$) & \\
\hline
     0-825& $-$0.035 & $-$0.081$\pm$0.029 & 2679 & $-$0.023 &~$-$0.060$\pm$0.012 &3153\\
\hline
     0-500&$-$0.017 &$-$0.083$\pm$0.030 & 2491 &$-$0.014 &$-$0.063$\pm$0.011 &2935\\
   500-800&$-$0.081 &$-$0.048$\pm$0.037 &  183 &$-$0.100 &$-$0.028$\pm$0.057 & 212\\
  800-1000&$-$0.130 &  +0.112$\pm$0.059 &   22 &$-$0.027 &  +0.138$\pm$0.056 &  29\\
 1000-1500&$-$0.064 &  +0.114$\pm$0.140 &   18 &$-$0.072 &$-$0.034$\pm$0.137 &  23\\
$\geq$1500&   --    &   --              &    5 &   --    &       --          &   5\\
\hline
Total     &   --   &   --             & 2719 &   --   &   --             &3204\\
\hline
\end{tabular}
\end{table*}

\subsection{Population Types of the Sample Stars}

We adopted the procedure detailed \citet{Aketal2015} to separate the sample stars into different populations. Thus, stars with $Z_{max}\leq 825$ pc are assumed to be thin-disc stars, while those with $Z_{max}>825$ pc are categorised as thick disc or perhaps halo stars. The above procedure is based on the Monte Carlo simulations with a wide range of a set of Galactic model parameters; i.e. local space density of the thick disc $0\leq n_{TK} \leq 15\%$, and exponential scale height of the thin and thick discs 200 $\leq H_{TN}\leq$ 350 pc and 500 $\leq H_{TK}\leq$ 1500 pc, respectively. The numerical value $Z_{max}=825$ pc is in agreement with the those previously appeared in the related research, \citep[i.e.][]{Ohjaetal1999, Siegetal2002, Karaalietal2004, Biletal2006}; hence it is robust. We are aware that there is some discussion in the literature whether a distinct two-component disc is a good description of our Galaxy. In any case, stars close to the Galactic mid-plane belong to the thin disc defined in situ via the spatial distribution of the Galactic components. However, we do not exclude the possibility of contamination of the thick-disc stars at metallicities typical of the thin disc in this region. 
          
\section{Results}
\subsection{Metallicity Gradient for Dwarfs on Circular Orbits}
We restricted our sample with $S/N\geq40$ (7935 stars) to obtain the best quality data and applied the constraint related to the planar eccentricity, $e_p\leq0.10$, to limit the sample stars with circular orbits. This sample lie at the solar circle, $7\leq R_m\leq 9$ kpc. The number of stars for which $[M/H]$ metallicity were determined is 3204, while 2719 indicated iron abundance $[Fe/H]$. We estimated metallicity gradients for the two sub-samples for a set of $Z_{max}$ intervals, $0<Z_{max}\leq825$, $0<Z_{max}\leq500$, $500<Z_{max}\leq800$, $800<Z_{max}\leq1000$, and $1000<Z_{max}\leq1500$ pc, which are presented in Table 1. The reason for estimation gradients for two sub-samples is that the number of stars with available $[M/H]$ metallicity is larger than the stars with iron abundance (for which we expect more accurate result) while iron gradients give us the chance to compare our results with those previously appeared in research literature. The value $Z_{max}=800$ pc is close to 825 pc, which is stated as an upper limit for the thin disc in the vertical distance. 

The iron gradients for the intervals $0<Z_{max}\leq 500$ and $500<Z_{max}\leq 800$ pc are $d[Fe/H]/dR_m=-0.083\pm0.030$ and $d[Fe/H]/dR_m=-0.048\pm0.037$ dex~kpc$^{-1}$, while for the two later intervals they are $d[Fe/H]/dR_m=+0.112\pm0.059$ and $d[Fe/H]/dR_m=+0.114\pm0.140$ dex~kpc$^{-1}$. Radial iron gradients close to zero or positive values at relatively high vertical distances are common in related research literature. However, we should also note that number of stars which cover the mentioned $Z_{max}$ intervals are only 22 and 18, respectively. We could not estimate any iron gradient for the interval $Z_{max}>1500$ pc due to a sample size of only five stars in this interval. The range of the mean iron abundances in four $Z_{max}$ intervals in Table 1 is $-0.13<[Fe/H]<-0.01$ dex, which indicates that all the stars in these intervals are thin-disc stars. The radial iron gradient for the interval $0<Z_{max}\leq825$ pc, i.e. $d[Fe/H]/dR_m=-0.081\pm0.029$ dex~kpc$^{-1}$, is rather close to the one for $0<Z_{max}\leq500$ pc. 

A metallicity gradient distribution could be observed for $[M/H]$ similar to $[Fe/H]$ in Table 1. The radial metallicity gradients $d[M/H]/dR_m=-0.063\pm0.011$ dex~kpc$^{-1}$ and $d[M/H]/dR_m=-0.028\pm0.057$ dex~kpc$^{-1}$ are compatible with the iron gradients in the same vertical distance intervals, $0<Z_{max}\leq500$ and $500<Z_{max}\leq800$ pc. The same case holds true also for the third $Z_{max}$ interval, $800<Z_{max}\leq 1000$ pc. However, the error for the metallicity gradient for the last $Z_{max}$ interval is large, $\pm$0.137. Therefore, there would be no meaning in comparing the metallicity gradient in this interval with the one corresponding to the iron metallicity, whose error is also large at $\pm$0.140. We should note that the large gradient errors are due to the small number of stars in those samples. The uncertainties in the derived gradients were determined via regression analysis. The metallicity gradients for two sub-samples are plotted in Fig. 4. 

We used both iron and metallicity abundances in deriving gradients in our study. Thus, we were able to compare the gradients estimated here with those previously appearing in the literature for either type of abundance ($d[Fe/H]/dR$ or $d[M/H]/dR$). All abundances (iron or metallicity) are taken from the same source, the RAVE DR4 database. The metallicity gradients have the advantage of a larger corresponding number of stars than those used for the iron gradients: 3204 and 2719, respectively. However, as one can see in Table 1 and Table 5, the iron gradients are (absolutely) larger than the metallicity ones. The iron originates in binary couples via the SN Ia process in a longer timescale, ($t\leq 8.5$ Gyr), meanwhile, the metallicity symbol ``M'' indicates the combination of a large number of metals including the $\alpha$-elements whose timescale is, $5.5\leq t\leq 10$ Gyr \citep{Brook12}, as well as their formation process, SN II, are different than the former ones. Calcium element (Ca) which dominates the overall metal abundance in RAVE survey is an $\alpha$-process product (although some Ca element could be produced by SN Ia as stated by \citet{Chiappini01}). Hence different gradients between iron and metallicity abundances is due to these differences just mentioned. 

\begin{figure*}
\centering
\includegraphics[scale=0.35, angle=0]{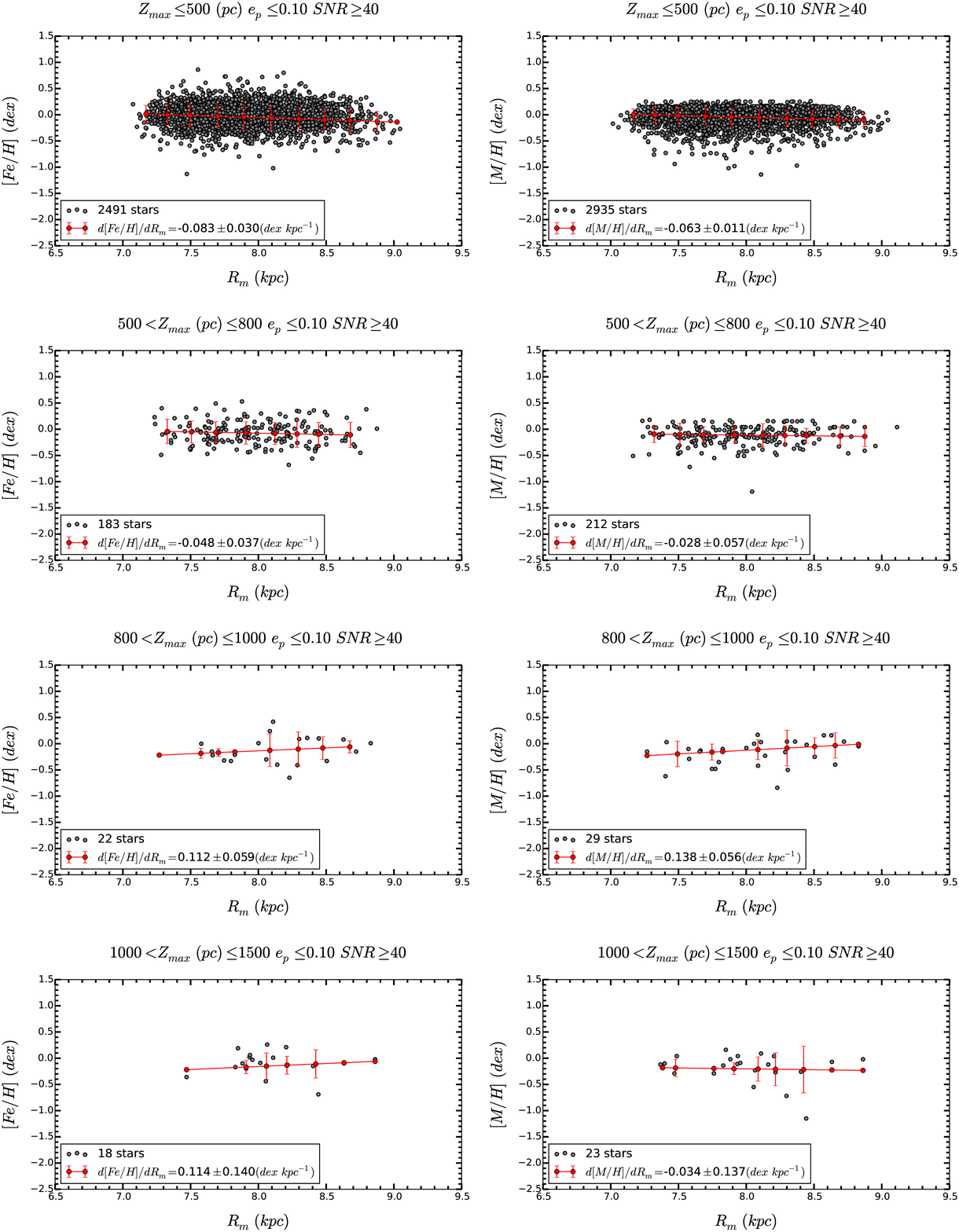}
\caption{Radial iron and metallicity distributions (left and right panels, respectively) for the F and G type dwarfs on circular orbits, $e_p\leq0.10$, or different $Z_{max}$ intervals.} 
\end {figure*}

\begin{figure*}
\centering
\includegraphics[scale=0.35, angle=0]{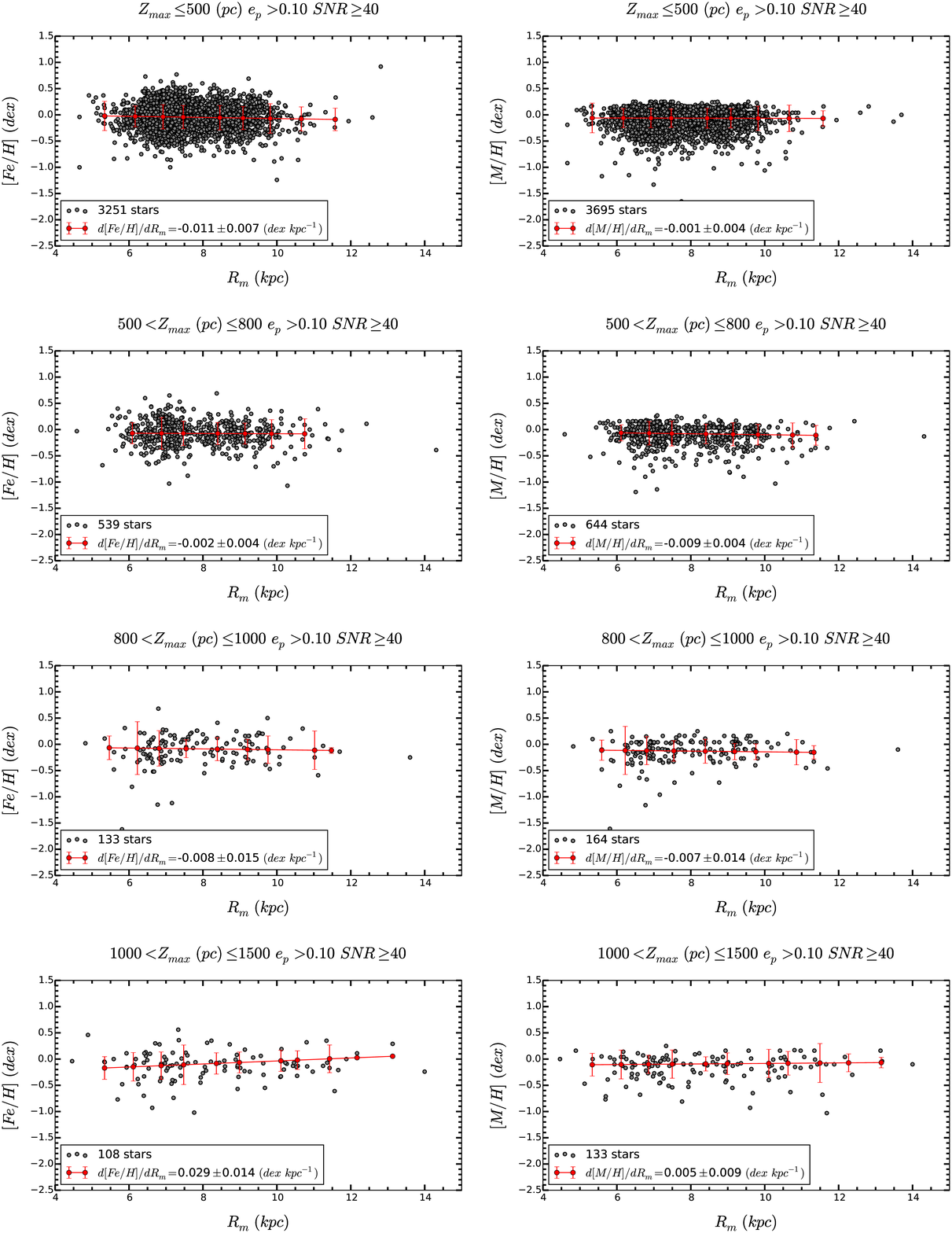}
\caption{Radial iron and metallicity distributions (left and right panels, respectively) for the F and G type dwarfs on elongated orbits, $e_p>0.10$, for different $Z_{max}$ intervals.} 
\end {figure*} 

\subsection{Metallicity Gradient for Dwarfs on Elongated Orbits}
We chose the dwarfs with planar eccentricities $e_p>0.10$ in the final sample (14361 stars) and limited them with $S/N\geq40$ to obtain an alternative sub-sample for investigation of a metallicity gradient. The number of stars on elongated orbits used for estimation of the $d[Fe/H]/dR_m$ and $d[M/H]/dR_m$ gradients are 4107 and 4731, respectively (Table 2). These stars occupy a larger radial distance interval than the sub-sample in Section 3.1; i.e. $5<R_m<12$ kpc. We estimated radial metallicity gradients for the same $Z_{max}$ intervals given in the preceding section. The $d[Fe/H]/dR_m$ iron gradients and the $d[M/H]/dR_m$ metallicity gradients for the three consecutive $Z_{max}$ intervals, $0<Z_{max}\leq500$, $500<Z_{max}\leq800$ and $800<Z_{max}\leq1000$ pc, are negative and close to zero, $\sim -0.01$ dex~kpc$^{-1}$, while both gradients for the interval $1000<Z_{max}\leq1500$ pc are small positive numbers. The iron and metallicity abundances for the stars in this sub-sample, $-0.11<[Fe/H]<-0.04$ and $-0.08<[M/H]<-0.03$ dex correspond to the thin-disc population at least for those at small $Z_{max}$ distances (i.e. $0<Z_{max}<800$ pc) where their number is much larger than those at larger $Z_{max}$ distances.

\begin{table*}[!htb]
\centering
\caption{Radial iron and metallicity gradients for the F and G type dwarfs on elongated orbits, $e_p>0.10$, for different $Z_{max}$ intervals. $N$ indicates the number of stars. Signal to noise ratio is $S/N\geq40$. The gradients are not significant (see the text).}
\begin{tabular}{ccccccc}
\hline
$Z_{max}$ range&$<[Fe/H]>$&$d[Fe/H]/dR_m$ &$N$&$<[M/H]>$&$d[M/H]/dR_m$ &$N$\\
\hline
      (pc)&(dex)&(dex~kpc$^{-1}$) & &(dex)& (dex~kpc$^{-1}$) & \\
\hline
     0-500&$-$0.045 &$-$0.011$\pm$0.007 & 3251 &$-$0.032 &$-$0.001$\pm$0.004&3695\\
   500-800&$-$0.066 &$-$0.002$\pm$0.004 &  539 &$-$0.055 &$-$0.009$\pm$0.004& 644\\
  800-1000&$-$0.062 &$-$0.008$\pm$0.015 &  133 &$-$0.080 &$-$0.007$\pm$0.014& 164\\
 1000-1500&$-$0.108 &  +0.029$\pm$0.014 &  108 &$-$0.070 &  +0.005$\pm$0.009& 133\\
$\geq$1500&   --  &          --         &   76 &   --    &    --            &  95\\
\hline
Total     &   --  &    --           & 4107 &  --   &    --            &4731\\
\hline
\end{tabular}
\end{table*}

The results are plotted in Fig. 5. There is a noticeable deficiency-slit of stars at the radial distance $R_m \sim 8$ kpc, which is the solar distance where the stars with circular orbits are concentrated (Section 3.1). We separated the stars subject in this section into two categories based on their radial distances and investigated metallicity gradients for both categories. The first category consists of stars with $R_m\leq8$ kpc, while the second one covers the stars with $R_m>8$ kpc. The results for the first category are presented in Table 3 and Fig. 6. All the iron and metallicity gradients are small positive or negative numbers. The positive gradients for small $Z_{max}$ distances from the Galactic plane have not appeared in related research literature until now. The case is similar for the iron and metallicity gradients estimated for the stars with $R_m>8$ kpc on elongated orbits (Table 4 and Fig. 7). That is, the insignificant iron and metallicity gradients for stars on elongated orbits are not radial distance dependent.     

\begin{table*}[!htb]
\centering
\caption{Radial iron and metallicity gradients for the F and G type dwarfs with radial distances $R_m\leq8$ kpc, on elongated orbits, $e_p>0.10$ (a sub-sample of stars in Table 2). $N$ indicates the number of stars. Signal to noise ratio is $S/N\geq40$.}
\begin{tabular}{ccccccc}
\hline
$Z_{max}$ range&$<[Fe/H]>$&$d[Fe/H]/dR_m$ &$N$&$<[M/H]>$&$d[M/H]/dR_m$ &$N$\\
\hline
      (pc)&(dex)&(dex~kpc$^{-1}$) & &(dex)& (dex~kpc$^{-1}$) & \\
\hline
     0-500&$-$0.027 &  +0.010$\pm$0.023 & 2275 &$-$0.035 &$-$0.004$\pm$0.018 &2564\\
   500-800&$-$0.071 &  +0.049$\pm$0.061 &  317 &$-$0.055 &  +0.033$\pm$0.044 & 373\\
  800-1000&$-$0.020 &  +0.056$\pm$0.075 &   73 &$-$0.097 &  +0.014$\pm$0.061 &  91\\
 1000-1500&$-$0.108 &$-$0.010$\pm$0.057 &   57 &$-$0.082 &$-$0.021$\pm$0.051 &  70\\
$\geq$1500&  --   &--                 &   38 &   --  &    --            &  48\\
\hline
Total     &    -- & --               & 2760 &   --  &   --             &3146\\
\hline
\end{tabular}
\end{table*}

\begin{table*}[!htb]
\centering
\caption{Radial iron and metallicity gradients for the F and G type dwarfs with radial distances $R_m>8$ kpc, on elongated orbits, $e_p>0.10$ (the complement sub-sample of the one in Table 3). $N$ indicates the number of stars. Signal to noise ratio is $S/N\geq40$.}
\begin{tabular}{ccccccc}
\hline
$Z_{max}$ range&$<[Fe/H]>$&$d[Fe/H]/dR_m$ &$N$&$<[M/H]>$&$d[M/H]/dR_m$ &$N$\\
\hline
      (pc)&(dex)&(dex~kpc$^{-1}$) & &(dex)& (dex~kpc$^{-1}$) & \\
\hline
     0-500 &$-$0.045 &$-$0.017$\pm$0.020 & 976 &$-$0.053 &   +0.011$\pm$0.014 & 1131\\
   500-800 &$-$0.070 &  +0.006$\pm$0.031 & 222 &$-$0.047 & $-$0.001$\pm$0.017 &  271\\
  800-1000 &$-$0.100 &  +0.007$\pm$0.052 &  60 &$-$0.057 & $-$0.021$\pm$0.030 &   73\\
 1000-1500 &$-$0.126 &  +0.004$\pm$0.023 &  51 &$-$0.086 &   +0.002$\pm$0.024 &   63\\
$\geq$1500 &    --   & --                &  38 &      -- &        --          &   47\\
\hline
Total      &     --&          --      &1347 &   --  &        --        &1585\\
\hline
\end{tabular}
\end{table*}

\begin{figure*}
\centering
\includegraphics[scale=0.35, angle=0]{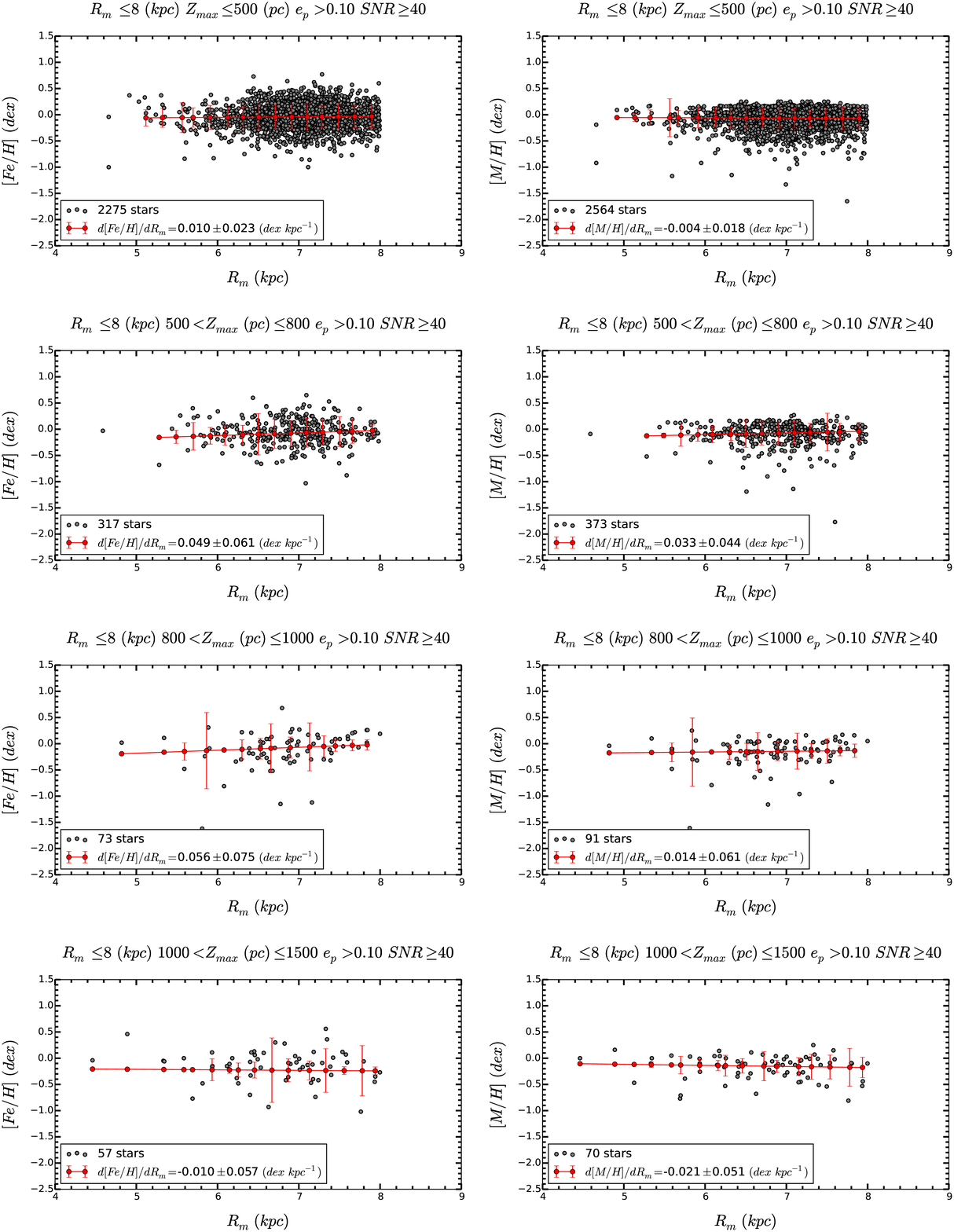}
\caption{Radial iron and metallicity distributions (left and right panels, respectively) for the F and G type dwarfs with radial distances $R_m\leq 8$ kpc on elongated orbits, $e_p>0.10$, for different $Z_{max}$ intervals.} 
\end {figure*} 

\begin{figure*}
\centering
\includegraphics[scale=0.35, angle=0]{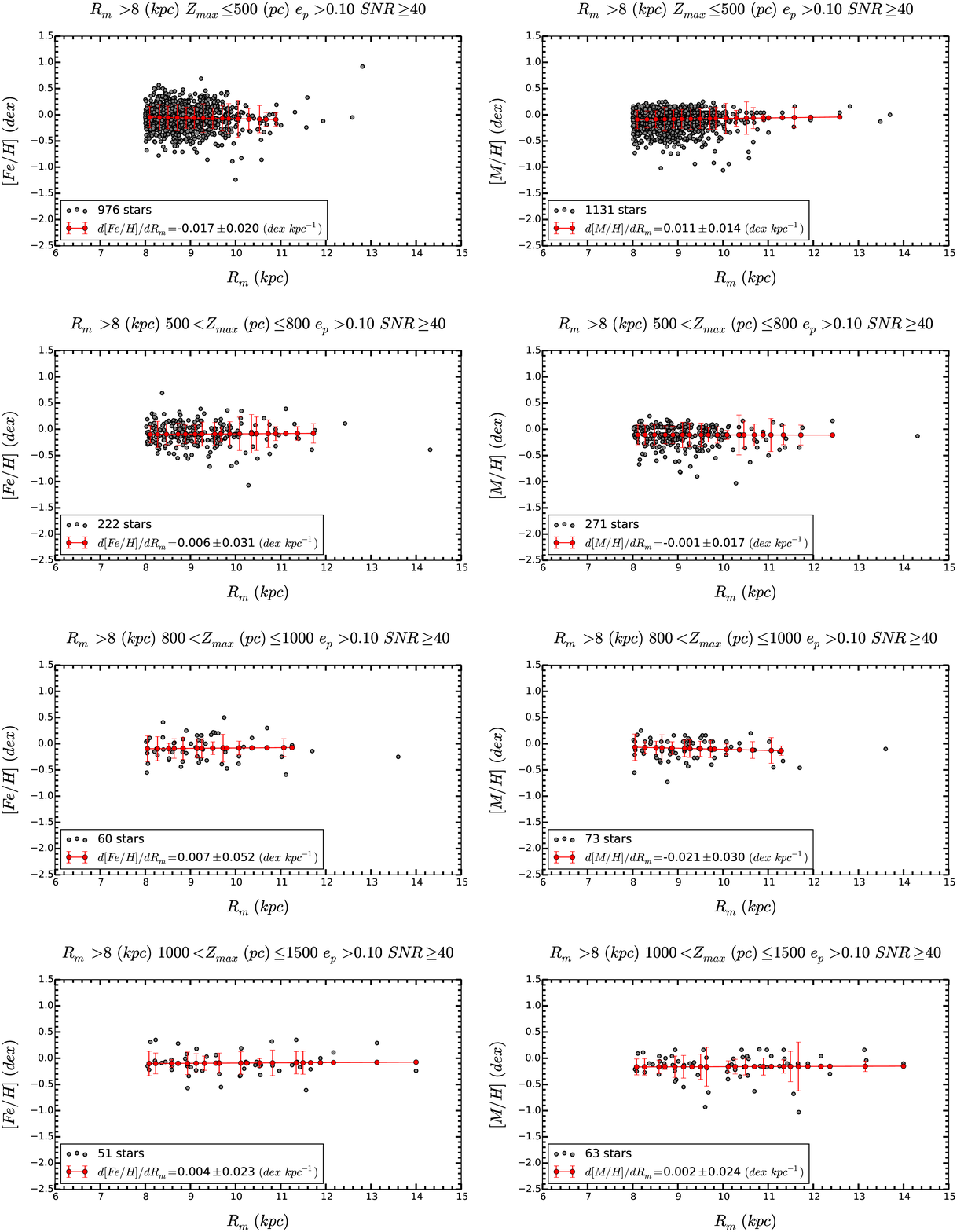}
\caption{Radial iron and metallicity distributions (left and right panels, respectively) for the F and G type dwarfs with radial distances $R_m>8$ kpc on elongated orbits, $e_p>0.10$, for different $Z_{max}$ intervals.} 
\end {figure*} 

\subsection{Metallicity Gradient for stars in the Vertical Direction}
We also estimated vertical metallicity gradients for both $[Fe/H]$ and $[M/H]$. We restricted our sub-sample to stars on circular orbits and $S/N\geq40$. The results are displayed in Fig. 8 and Fig. 9, for the iron and metallicity abundances, respectively. The iron gradient, $d[Fe/H]/dZ_{max}$, is $-0.176\pm 0.039$ dex~kpc$^{-1}$ for $Z_{max}\leq 825$ pc (upper panel) while it is $-0.119\pm 0.036$ dex~kpc$^{-1}$ for $Z_{max}\leq 1500$ pc (lower panel). The metallicity gradients, $d[M/H]/dZ_{max}$, in Fig. 9 are compatible with the corresponding ones in Fig. 8, i.e. $-0.155\pm 0.035$ dex~kpc$^{-1}$ for $Z_{max}\leq 825$ pc and $-0.126\pm 0.048$ dex~kpc$^{-1}$ for $Z_{max}\leq 1500$ pc. There is a noticeable flattening of the vertical gradient with the height above the Galactic mid-plane, which is a common result in previous research literature.  

\section{Discussion}

We estimated iron and metallicity gradients in the radial and vertical directions with F and G type dwarfs selected from the recent RAVE DR4 database \citep{Kordopatis13}. We restricted the data with $S/N\geq40$ to obtain the probable best results. We also applied two constraints to obtain a limited sample of F and G dwarfs and estimated iron and metallicity gradients for the stars in this sample. The constraints are as follows: $i$) $Z_{max}\leq 825$ pc and $ii$) $e_p\leq 0.10$. The first parameter was first used by \citet{Aketal2015}, while the second appeared in \citet{Huang15}, who used the range $e_p<0.13$ to separate the cold thin-disc stars from the thick-disc stars. The vertical distance $Z_{max}=825$ pc is the place where space densities of the thin and thick discs are compatible. The second parameter, $e_p=0.10$, is the mode of $e_p$ estimated in our sample. That is, our constraints define a thin disc with F and G type dwarfs on circular orbits. If some thick-disc stars contaminate our sample, the resulting metallicity distribution will not be different than a pure thin-disc metallicity distribution. Because, the metallicity of the thick disc stars in question would be compatible to that of the thin-disc stars. We used the stars with $Z_{max}\leq 825$ pc and $e_p\leq 0.10$ to estimate iron and metallicity gradients with respect to $R_m$ radius. However, we also estimated radial iron and metallicity gradients beyond the vertical distance $Z_{max}=825$ pc for comparison purposes. 

We adopted the vertical distance intervals as $0<Z_{max}\leq500$, $500<Z_{max}\leq 800$, $800<Z_{max}\leq 1000$, and $1000<Z_{max}\leq1500$ pc in our calculations. Thus, we decreased the $Z_{max}$ distance in the first constraint by 25 pc. The iron and metallicity gradients for the intervals $0< Z_{max}\leq500$, $500<Z_{max}\leq800$ pc, for the most likely thin disc thus defined, are compatible with each other and they are at the level of the radial iron gradients previously appearing in related research. Actually $d[Fe/H]/dR_m=-0.083\pm0.030$ and $d[Fe/H]/dR_m=-0.048\pm0.037$ dex~kpc$^{-1}$ in our study (Table 1) are generally compatible with the ones in \citet{Boeche13}, i.e. $d[Fe/H]/dR_g=-0.059\pm0.005$ dex~kpc$^{-1}$, estimated for the interval $0.4<Z_{max}\leq0.8$ kpc, and in \citet{Cheng12}, $d[Fe/H]/dR=-0.055$ dex~kpc$^{-1}$, estimated for the interval $0.25<|Z|\leq0.50$ kpc, where $R_g$ and $R$ are the guiding and Galactocentric radii, respectively. The range of the mean iron abundances for the dwarfs in question is $-0.064\leq[Fe/H]\leq -0.017$ dex which indicates that dwarfs subject to the aforementioned radial iron (metallicity) gradient are thin disc stars. The Toomre diagram in Fig. 10 confirms our argument kinematically. The $V_{LSR}$ and $\sqrt{U^{2}_{LSR}+W^{2}_{LSR}}$ velocity components of the stars in question lie in a circle with radius 50 km~s$^{-1}$. That is, F and G type dwarfs with $Z_{max}\leq 800$ pc on circular orbits are thin-disc stars and exhibit a notable radial metallicity gradient. The radial distance range of these stars is $7\leq R_m \leq 9$ kpc, indicating that they lie at the solar circle. 

The iron (and metallicity) gradients for F and G dwarfs in the intervals $800<Z_{max}\leq 1000$ and $1000<Z_{max}\leq 1500$ pc consist of three small positive values, $d[Fe/H]/dR_m=+0.112\pm0.059$, $d[Fe/H]/dR_m=+0.114\pm0.140$ and $d[M/H]/dR_m=+0.138\pm0.056$ dex~kpc$^{-1}$ (Table 1), in agreement with the corresponding values in the literature \citep[cf.][]{Cheng12,Boeche13}, along with one negative value with large error, $d[M/H]/dR_m=-0.034\pm0.137$ dex~kpc$^{-1}$. We should note that the number of stars corresponding with each gradient just cited is not greater than 30. According to the arguments in the literature, zero or even positive radial gradients are due to the effect of the thick-disc stars at relatively large vertical distances. The metallicity range of these stars is almost the same as the range for the stars with $0<Z_{max}\leq800$ pc, as investigated in the preceding paragraph. That is, they should be thin disc stars, if we use only the metallicity parameter to determine their population types. However, their velocity components are a bit different than the ones discussed in the preceding paragraph. These stars lie at the outermost part of the velocity circle with a radius of 50 km~s$^{-1}$ (Fig. 10). 

\begin{figure}[h]
\centering
\includegraphics[trim=0cm 0cm 0cm 0cm, clip=true, scale=0.35]{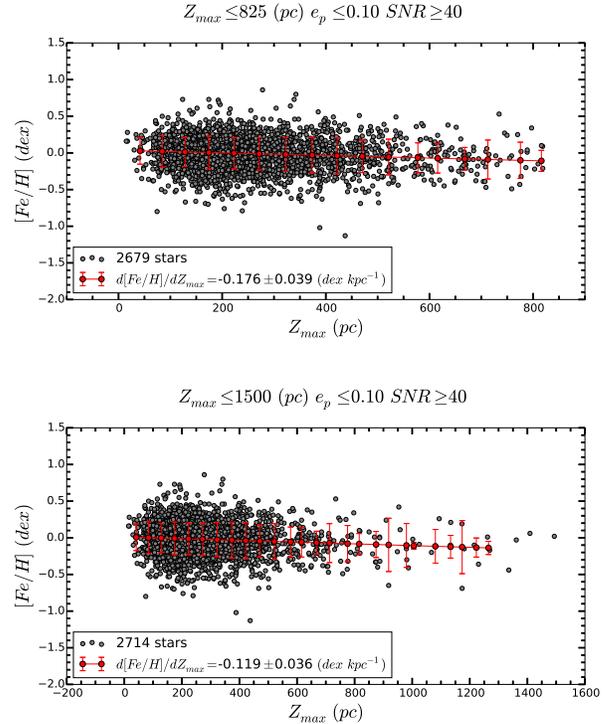}
\caption{Vertical iron distributions for two distance intervals: $Z_{max}\leq 825$ pc (upper panel) and $Z_{max}\leq 1500$ pc (lower panel).} 
\end {figure} 

\begin{figure}[h]
\centering
\includegraphics[trim=0cm 0cm 0cm 0cm, clip=true, scale=0.35]{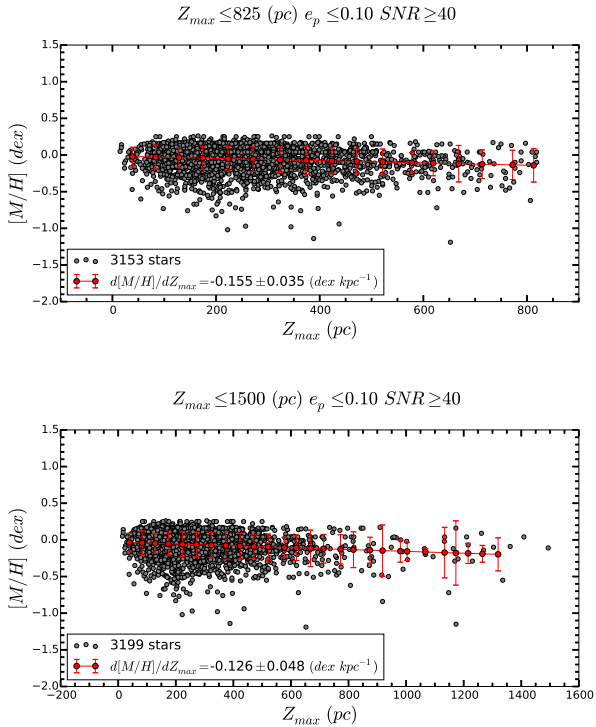}
\caption{Vertical metallicity distributions for two distance intervals: $Z_{max}\leq 825$ pc (upper panel) and $Z_{max}\leq 1500$ pc (lower panel).} 
\end {figure} 

\begin{figure}[h!]
\centering
\includegraphics[trim=0cm 3cm 0cm 3cm, clip=true, scale=0.35]{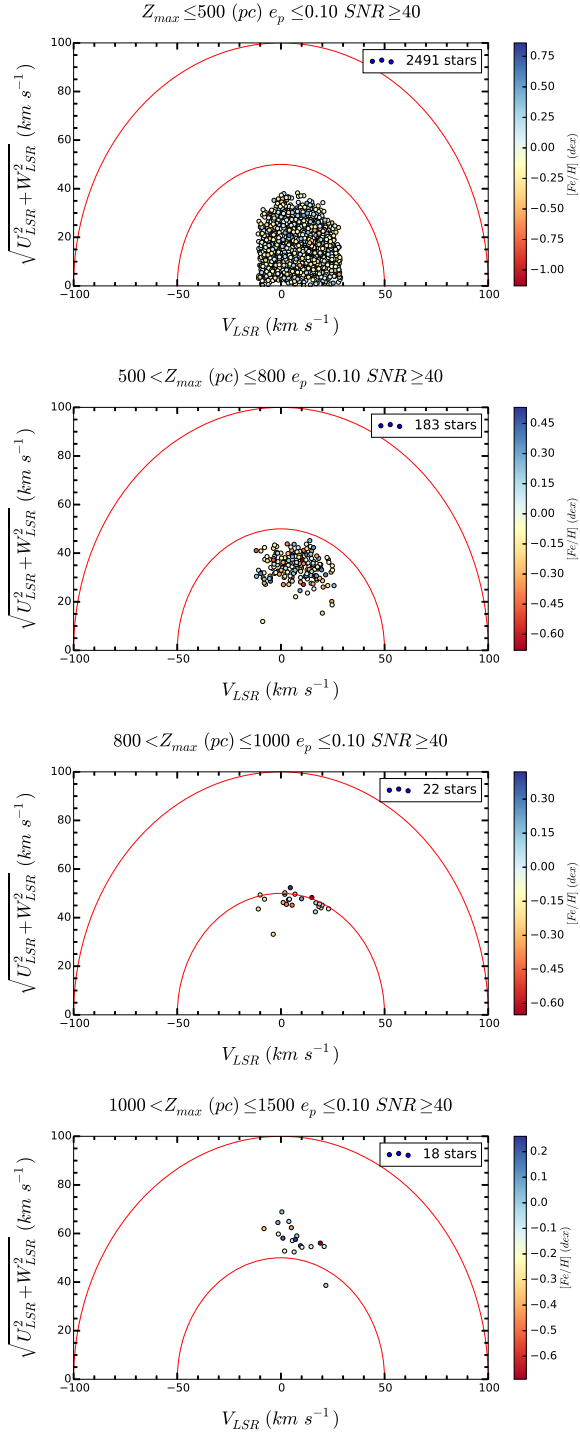}
\caption{Toomre diagram for the F and G type dwarfs on circular orbits, $e_p\leq 0.10$, for different $Z_{max}$ intervals: $0<Z_{max}\leq500$ pc, $500<Z_{max}\leq 800$ pc, $800<Z_{max}\leq 1000$ pc, and $1000<Z_{max}\leq 1500$ pc.} 
\end {figure}

\begin{table*}[!htb]
\centering
\caption{Radial iron and metallicity gradients with $R_g$ radii for the F and G type dwarfs on circular orbits, $e_p\leq0.10$, for different $Z_{max}$ intervals. $N$ indicates the number of stars. Signal to noise ratio is $S/N\geq40$.}
\begin{tabular}{ccccccc}
\hline
$Z_{max}$ range&$<[Fe/H]>$&$d[Fe/H]/dR_g$ &$N$&$<[M/H]>$&$d[M/H]/dR_g$ &$N$\\
\hline
      (pc)&(dex)&(dex~kpc$^{-1}$) & &(dex)& (dex~kpc$^{-1}$) & \\
\hline
     0-825&$-$0.035 &$-$0.074$\pm$0.028 & 2679 &$-$0.023 &$-$0.050$\pm$0.017 &3153\\
\hline
     0-500&$-$0.017 &$-$0.083$\pm$0.030 & 2491 &$-$0.014 &$-$0.062$\pm$0.018& 2935\\
   500-800&$-$0.081 &$-$0.065$\pm$0.039 &  183 &$-$0.100 &$-$0.055$\pm$0.045&  212\\
  800-1000&$-$0.130 &  +0.194$\pm$0.084 &   22 &$-$0.027 &  +0.154$\pm$0.041&   29\\
 1000-1500&$-$0.064 &  +0.126$\pm$0.113 &   18 &$-$0.072 &$-$0.080$\pm$0.160&   23\\
$\geq$1500&   --    &          --       &    5 &   --    &    --            &    5\\
\hline
Total     &   --    &    --             & 2719 &  --   &    --            &3204\\
\hline
\end{tabular}
\end{table*}

We also investigated the iron and metallicity gradients for the F and G type dwarfs on elongated orbits, $e_p>0.10$. Again, we chose the data with $S/N\geq40$ to obtain the probable best results. Significant metallicity gradients could not be derived for four $Z_{max}$ intervals for this sample (Table 2 and Fig. 5). The range of the mean iron and metal abundances of the stars in this sample are $-0.11<[Fe/H]<-0.04$ and $-0.08\leq[M/H]<-0.03$ dex, respectively. Stars with these metallicities and short $Z_{max}$ distances probably belong to the thin-disc population. However, the metal-poor ends of these abundance intervals may include some thick-disc stars. We separated this sample into two sub-samples by means of their radial distance, $R_m\leq8$ and $R_m>8$ kpc, and estimated iron and metallicity gradients for each sub-sample. Neither for the sub-sample stars with $R_m\leq 8$ kpc, nor for those with $R_m>8$ kpc could any significant iron or metallicity gradient be detected. This indicates that the insignificance for the F and G type dwarfs on elongated orbits is due to planar eccentricity, and that radial distance does not have any constraint on such small (positive or negative) gradients.      
        
We compared the Toomre diagrams of the two sub-samples to reveal any differences in kinematics. Fig. 11 shows that space velocity components for these two sub-samples are much larger than the F and G type dwarfs on the circular orbits (Fig. 10). However, there are some noticeable differences between the space velocity components of stars in the two sub-samples as well: the range of the velocity components and the absolute value of a given space velocity component for stars with $R_m\leq8$ kpc are both larger than the corresponding values for stars with $R_m>8$ kpc. The diagrams in Fig. 11 indicate that both sub-samples are probably a mixture of thin- and thick-disc stars with different percentages. However, as stated above, the range of the iron and metallicity abundances of the stars in these two sub-samples favors a sub-sample of thin-disc population. The solution to this conflict is to assume that the thick-disc stars in two sub-samples should be the metal-rich tail of the thick-disc population.   

The vertical iron gradient $d[Fe/H]/dZ_{max}=-0.176\pm0.039$ dex~kpc$^{-1}$ in our work is compatible with the one found by \citet{Boeche14}, $d[Fe/H]/dZ_{max}=-0.112\pm0.007$ dex~kpc$^{-1}$, estimated for the red clump stars with $z\leq2$ kpc. Our vertical metallicity gradient, $d[M/H]/dZ_{max}=-0.155\pm0.035$ dex~kpc$^{-1}$ is also compatible with the corresponding values appearing in the literature which are estimated for stars with different vertical distances and Galactic latitudes; i.e. $d[M/H]/dz=-0.16\pm0.02$ dex~kpc$^{-1}$ for $z\leq 3$ kpc and $b=+45^{\circ}$, \citep{Ak07a}; $d[M/H]/dz=-0.22\pm0.03$ dex~kpc$^{-1}$ for $z\leq 3$ kpc and $b=+60^{\circ}$, \citep{Ak07b}; $d[M/H]/dZ_{max}=-0.167\pm0.011$ dex~kpc$^{-1}$ for $Z_{max}\leq 1.5$ kpc, \citep{Bilir12}.

In recent years, metallicity gradient estimations have been based on the guiding radius $R_g$ \citep[cf.][]{Boeche13,Boeche14}. In our study, we estimated the metallicity gradients using mean radius $R_m$ and compared our results with the ones estimated with $R_g$ in previous research. We calculated the metallicity gradients of the sub-samples in Table 1 according to the guiding radius $R_g$. Results are listed in Table 5. There are clear negative metallicity and iron gradients for the $Z_{max}$ intervals below 800 pc, while there are positive gradients above this $Z_{max}$ value. It is clear that the results in Tables 1 and 5 are in good agreement. Thus, we conclude that both $R_g$ and $R_m$ can be used in the estimation of the radial metallicity gradients. 

Small differences in the two different studies are always expected due to different types of radii such as mean radius ($R_m$) and guiding radius ($R_g$), different Galactic potential and different procedures for distance estimation. In our study, we used $R_m$ distances, the recent {\it MWpotential2014} code of \citet{Bovy15}, a different procedure for distance estimation \citep{Bilir08, Bilir09}, and another for identifying the F and G type dwarfs. Despite these differences, we obtained iron and metallicity gradients compatible with those in previously-published research. 

Another agreement holds between the iron and metallicity gradients estimated in \citet{Anders14} with respect to two different radial distances; i.e. guiding radius $R_g$ (their Table 3) and Galactocentric radius $R$ (their Fig. A.1, upper left panel), and our iron and metallicity gradients based on the data of stars on circular orbits (Table 1). It is interesting that the metallicity gradient with respect to the Galactocentric radius in \citet{Anders14}, $d[M/H]/dR=-0.082\pm 0.002$ dex~kpc$^{-1}$ is rather close to the iron gradient with respect to the $R_m$ radius, $d[Fe/H]/dR_m=-0.083\pm0.030$ dex~kpc$^{-1}$ in our study. Different gradients cited above can be attributed to different star categories in the Galactic disc. For example different components of the disc, such as thin disc and thick disc, could have different gradients and even gradients in a single component could be a function of age. It is most likely that the lower gradient observed in stars of higher eccentricity corresponds to a flatter gradient for older stars. 

We mentioned in the introduction that there is a bias against metal-rich stars if one uses $R_m$ distances; i.e. only stars with large orbital eccentricities among a sample with small $R_m$ distances can reach the solar circle. But such stars are metal-poor relative to the ones on circular orbits with the same $R_m$ radial distances which can not reach to the solar circle. Thus, the number of metal-rich stars at the solar circle decreases. However, in our case stars with large orbital eccentricities were removed by applying the constraint $e_p\leq 0.1$. Hence, the effects of the bias in question have been minimized.

\begin{figure*}
\centering
\includegraphics[trim=3cm 1cm 2cm 1cm, clip=true, scale=0.40]{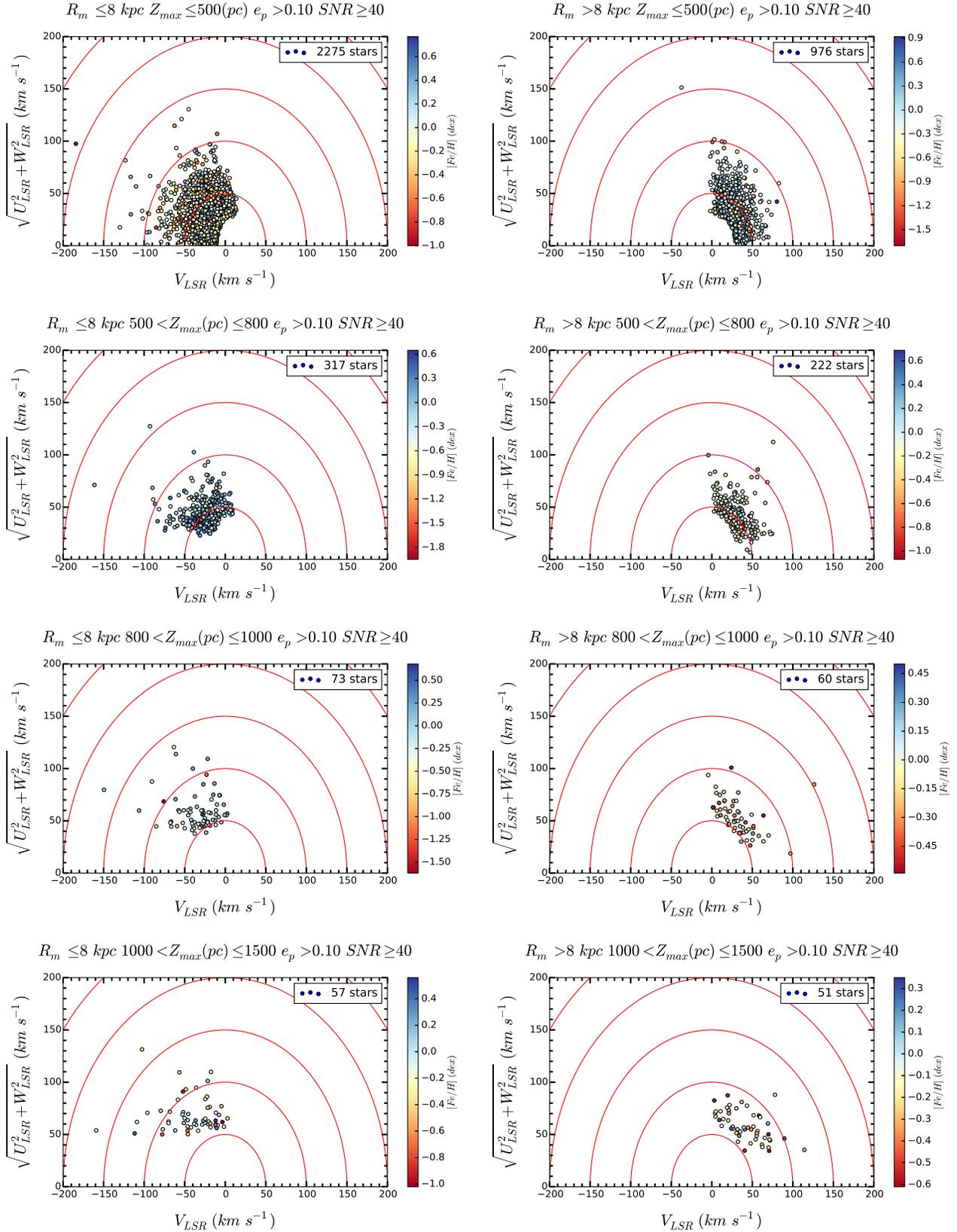}
\caption{Toomre diagram for the F and G type dwarfs with radial distances $R_m\leq8$ and $R_m>8$ kpc on elongated orbits, $e_p>0.10$, for different $Z_{max}$ intervals: $0<Z_{max}\leq500$ pc, $500<Z_{max}\leq 800$ pc, $800<Z_{max}\leq 1000$ pc, and $1000<Z_{max}\leq 1500$ pc.} 
\end {figure*} 

{\bf Conclusion:} Two constraints, $Z_{max}\leq825$ pc and $e_p\leq0.1$, provide a sample of thin-disc stars at the solar circle with radial iron and metallicity gradients in the vertical distance intervals, $0<Z_{max}\leq500$ and $500<Z_{max}\leq800$ pc. Negative or positive small gradients at further vertical distances probably originate from the mixture of two Galactic disc populations; i.e. thin and thick discs. The metallicity gradients could not be detected for stars on elongated orbits, $e_p>0.1$. Vertical iron and metallicity gradients estimated for the thin-disc sample are compatible with the ones in previous research.

\section{Acknowledgments}
Authors are grateful to the anonymous referee for his/her considerable contributions to improve the paper. The authors are very grateful to Mrs. Shannon Darling Oflaz for her careful and meticulous reading of the English manuscript. This study has been supported in part by the Scientific and Technological Research Council (T\"UB\.ITAK) 114F347. Part of this work was supported by the Research Fund of the University of Istanbul, Project Numbers: 39170, 39742, and 48027.

\end{document}